\documentclass[aps,pre,twocolumn,showpacs]{revtex4}
\usepackage{amssymb}
\usepackage{amsmath}
\usepackage{mathptmx}
\usepackage{graphicx}
\usepackage{bm}

\begin{document}

\title{Spin induced nonlinearities in the electron MHD regime}
\author{Martin Stefan}
\author{Gert Brodin}
\author{Mattias Marklund}

\begin{abstract}
We consider the influence of the electron spin on the nonlinear propagation
of whistler waves. For this purpose a recently developed electron two-fluid
model, where the spin up- and down populations are treated as different
fluids, is adapted to the electron MHD regime. We then derive a nonlinear
Schr\"{o}dinger equation for whistler waves, and compare the coefficients of
nonlinearity with and without spin effects. The relative importance of spin
effects depend on the plasma density and temperature as well as the external
magnetic field strength and the wave frequency. The significance of our
results to various plasmas are discussed.
\end{abstract}

\affiliation{Department of Physics, Ume{\aa } University, SE--901 87 Ume{\aa }, Sweden}
\pacs{52.27.-h, 52.27.Gr, 67.57.Lm}
\maketitle


\section{Introduction}


The dynamics of dense ionized matter, cold matter in strong magnetic fields,
and nonlinear degenerate plasmas has applications to both laboratory and
naturally occurring systems. Many such matter states goes under the
collective notation of quantum plasmas. A multitude of studies devoted to
quantum plasma effects can be found in the literature, much of it inspired
by works like e.g. Refs.\ \cite{Bohm-Pines,Pines-1953}. In the last decade
there has been somewhat of a surge in the interest of quantum plasmas \cite%
{Haas-2000,Manfredi-review,Garcia-2005,Shukla-Stenflo-2006,Shukla-Eliasson-2006,Marklund-2007,Brodin-2007,Shukla-2007,Brodin-2008,g-factor}%
. Interesting applications of this field can be found in for example
plasmonics \cite{Atwater-Plasmonics,Marklund-EPL-plasmonics}, quantum wells 
\cite{Manfredi-quantum-well} and ultracold plasmas \cite{Ultracold}. Common
to such applications are rather "extreme" parameters, compared to most
laboratory and space plasmas. More specifically, the plasma densities needs
to be very high and/or the temperatures correspondingly low. For
astrophysical plasmas, the situation is somewhat different since the strong
magnetic fields \cite{Astro,Melrose-book} may induce various types of
quantum effects. In most of the above studies, the spin effects plays little
or no dynamic role. The inclusion of collective spin dynamics \cite%
{Cowley-1986,Kulsrud-1986}gives rise to new modes in plasma, both at fluid 
\cite{Brodin-2007,NJP1} and kinetic scale \cite{g-factor}. Indeed, even
dusty plasmas can show interesting magnetization effects \cite{NJP2,AIPConf}%
. In Ref.\ \cite{Brodin-2008} the picture outlined above, concerning the
necessary parameter space for quantum effects to be important, was to some
extent modified, as it was shown that the spin properties of electrons can
be important in plasmas even outside the high density/low temperature
regime, also for moderate magnetic field strengths. Moreover, a recent focus
on the nonlinear regime in quantum plasmas \cite%
{nitin,shukla-ali-etal.,shukla-eliasson-review} makes the question of
magnetization nonlinearities interesting.

Motivated by the above, we will in the present work further extend the
analysis put forward in Ref.\ \cite{Brodin-2008}. In that work, the
electrons were described using a two-fluid model, where the spin-up and
spin-down populations relative to the magnetic field were treated as
different fluids in the standard magnetohydrodynamic (MHD) regime. Here we
will extend that treatment to cover the electron-MHD (EMHD) regime. In
particular we will study weakly nonlinear whistler waves, and derive a
nonlinear Schr\"{o}dinger (NLS) equation for the slowly varying amplitude,
both using classical weakly relativistic theory and the recent two-fluid
spin model, adapted for the EMHD regime. By comparing the nonlinear
coefficients in the different models and their dependence on the plasma
parameters (temperature, density external magnetic field strength), the
relative importance of the electron spin effects in various regimes can be
deduced. The result that electron spin effects can be important in other
regimes, as compared to certain much studied quantum effects [such as the
Bohm-de Broglie potential (see also Ref.\ \cite{Melrose-2009} for a
discussion) and the Fermi pressure], is confirmed. Finally we compare the
relative importance of electron spin effects in the EMHD regime with that in
the standard MHD regime.


\section{Two-fluid model with spin}


The purpose of our work is to compare the nonlinearities from classical and
quantum effects, respectively, in the EMHD regime. As a starting point, we
follow Ref.\ \cite{bulanov91} and derive the EMHD model. We assume that the
time-scale of interest is small enough that the ion motion, due to their
large mass compared to the electrons, can be neglected. We also neglect the
displacement current in Amp\`{e}res law. Furthermore by assuming an
isothermal pressure model and no dissipation, the governing equation can be
written as 
\begin{equation}
\frac{\partial }{\partial t}\left( \mathbf{B}-d_{e}^{2}\Delta \mathbf{B}%
\right) =-\alpha \nabla \times \left[ \left( \nabla \times \mathbf{B}\right)
\times \left( \mathbf{B}-d_{e}^{2}\Delta \mathbf{B}\right) \right] ,
\label{EMHD}
\end{equation}%
where $\mathbf{B}$ is the magnetic field, $d_{e}^{2}=c^{2}/\omega _{pe}^{2}$%
, $\omega _{pe}^{2}=q_{e}^{2}n/m_{e}\epsilon _{0}$ being the electron plasma
frequency, and $\alpha =1/nq_{e}\mu _{0}$. Studying the linear modes of this
equation propagating parallel to the magnetic field one finds whistler waves
with dispersion relation $\omega (k)=\omega _{ce}c^{2}k^{2}/\omega _{pe}^{2}$%
. However, since this model does not allow any density fluctuations, an
attempt to derive an NLS-equation shows that the model do not give rise to
any cubic nonlinearities. To still be able to do the intended comparison, we
still use the assumptions corresponding to the EMHD regime but perform a
more general treatment, allowing for relativistic particle velocities and
using a multifluid model that permits density perturbations.

The quantum model used in the comparison is obtained formally by starting
from the Pauli Hamiltonian as is done in Ref.\ \cite{Marklund-2007}, using
ensemble averaging to obtain the fluid equations 
\begin{eqnarray}
\frac{\partial }{\partial t}n+\nabla \cdot \left( n\mathbf{v}\right) &=&0 \\
m_{e}n\left( \frac{\partial }{\partial t}+\mathbf{v}\cdot \nabla \right) 
\mathbf{v} &=&q_{e}n\left( \mathbf{E}+\mathbf{v}\times \mathbf{B}\right)
-\nabla p+\mathbf{F}_{\text{spin}}
\end{eqnarray}%
where 
\begin{equation}
\mathbf{F}_{\text{spin}}=\pm \mu _{B}n\nabla B,
\end{equation}%
the sign depending on the spin orientation $\mathbf{S}_{\pm }=\pm ({n\hbar }/%
{2})\hat{\mathbf{B}}$ relative to the magnetic field. Thus, assuming a
two-electron fluid model, for which the distinction between the fluids is
through their relative spin orientation, we have 
\begin{eqnarray}
m_{e}n_{\pm }\frac{d\mathbf{v}_{\pm }}{dt} &=&n_{\pm }q_{e}\left( \mathbf{E}+%
\mathbf{v}_{\pm }\times \mathbf{B}\right) \pm \mu _{B}n_{\pm }\nabla
B-k_{B}T\nabla n_{\pm }.  \label{spinmomentum} \\
\frac{\partial }{\partial t}n_{\pm } &=&-\nabla \cdot \left( n_{\pm }\mathbf{%
v}_{\pm }\right) ,
\end{eqnarray}%
where the subscript $\pm $ denotes spin orientation parallel or
anti-parallel to the external magnetic field respectively, and $B=\left\vert 
\mathbf{B}\right\vert $. In case the up- and down spin populations are not
equal, there will be a net magnetization and a corresponding magnetization
current. Thus within this model the total current density to be used in
Ampere's law is written

\begin{equation}
\mathbf{j}=q_{e}(n_{+}\mathbf{v}_{+}+n_{-}\mathbf{v}_{-})+\mu _{B}\nabla
\times \left( n_{+}\hat{\mathbf{B}}-n_{-}\hat{\mathbf{B}}\right)
\label{current}
\end{equation}%
where the last term is the magnetization current due to the spin, and $\hat{%
\mathbf{B}}=\mathbf{B/}B$ is a unit vector in the direction of the magnetic
field.


\section{Linear theory}


Linearizing the momentum and fluid equations around a constant magnetic
field $\mathbf{B}_{0}=B_{0}\hat{\mathbf{z}}$, and assuming transversal waves
propagating parallel to this external magnetic field we can deduce 
\begin{equation}
\mathbf{v}=\hat{\sigma}\mathbf{E},
\end{equation}%
from Eq. (\ref{spinmomentum}), where 
\begin{equation*}
\hat{\sigma}=\frac{q_{s}}{m_{s}}\hat{M}^{-1}=\frac{q_{s}}{m_{s}\left( \omega
_{cs}^{2}-\omega ^{2}\right) }\left( 
\begin{array}{cc}
-i\omega & \omega _{cs} \\ 
-\omega _{cs} & -i\omega%
\end{array}%
\right)
\end{equation*}%
is the conductivity tensor, and $\omega _{cs}=q_{s}B_{0}/m_{s}$ is the
cyclotron frequency for particle species $s$, and we let the vectors here
just contain the parts perpendicular to $\hat{\mathbf{z}}$. Note that since
the variations of $B$ is nonlinear in the amplitude for parallel
propagation, the spin effects do not enter here. Furthermore, when the
thermal energy $k_{B}T$ is much larger than the energy difference between
the spin states, $\mu _{B}B_{0}$, the difference between the number density
of the spin-up and down populations, $n_{0+}-n_{0-}$, in the thermodynamic
ground state is exponentially small (proportional to $\exp (-\mu
_{B}B_{0}/k_{B}T)$), and hence we can omit the linearized part of the
magnetization current. Thus in this approximation no quantum effects remains
in the linearized theory. From Maxwell equations, we then obtain 
\begin{equation}
\hat{D}\mathbf{E}=\mathbf{0}  \label{de}
\end{equation}%
were 
\begin{equation}
\hat{D}=\left[ -\frac{k^{2}c^{2}}{\omega ^{2}}\hat{1}+\hat{1}-\sum_{s}\frac{%
\omega _{ps}^{2}}{\omega \left( \omega _{cs}^{2}-\omega ^{2}\right) }\left( 
\begin{array}{cc}
-\omega & -i\omega _{cs} \\ 
i\omega _{cs} & -\omega%
\end{array}%
\right) \right] .  \label{ddd}
\end{equation}%
Thus, we obtain the general dispersion relation 
\begin{equation}
\omega ^{2}=k^{2}c^{2}\left[ 1+\sum_{s}\frac{\omega _{ps}^{2}}{\omega
_{cs}^{2}-\omega ^{2}}\left( 1-\frac{\omega _{cs}}{\omega }\right) \right]
^{-1}
\end{equation}%
for this geometry

To obtain the EMHD limit, we disregard the displacement current in Amp\`{e}%
res law and regard the ions as fixed. This corresponds to neglecting the
unit matrix in Eq. (\ref{de}) and letting $\omega _{ce} \ll\omega $. By this
procedure we get the dispersion relation 
\begin{equation}  \label{emhd1}
\omega (k)=\frac{\omega _{ce}}{\omega _{pe}^{2}}c^{2}k^{2}=\frac{B_{0}k^{2}}{%
q_{e}n_{e}\mu _{0}}\equiv \alpha B_{0}k^{2}
\end{equation}
for small amplitude whistler waves propagating parallel to the external
magnetic field.

The dispersion relation (\ref{emhd1}) was obtained by starting from the
two-fluid model and then taking the EMHD limit. If one would start directly
from the EMHD plasma equation, the corresponding result would be 
\begin{equation}
\omega (k)=\frac{\alpha B_{0}k^{2}}{1+{c^{2}k^{2}}/{\omega _{p}^{2}}}.
\end{equation}%
This difference is due to the fact that for the EMHD assumptions to apply,
the parallel (to the external magnetic field) wavenumber $k_{\parallel }$
must obey $k_{\parallel }$ $\ll {\omega _{p}/c}$. \ Thus in our case with
parallel propagation (${k=}k_{\parallel }$) we must approximate the
denominator with unity, in which case the different expressions agree. As a
side note, for general directions of propagation the factor $(1+{c^{2}k^{2}}/%
{\omega _{p}^{2})}^{-1}$ appears correctly from EMHD theory, but since only
the perpendicular part of the wavenumber $k_{\perp }$ is allowed to be
comparable to the inverse skin depth (i.e. $k_{\perp }\sim {\omega _{p}/c}$)$%
\,$, we have $(1+{c^{2}k^{2}}/{\omega _{p}^{2})}^{-1}=(1+{c^{2}}k_{\perp
}^{2}/{\omega _{p}^{2})}^{-1}$in that case.

As usual, by using an ansatz of a weakly modulated amplitude in the EMHD
model Eq. (\ref{EMHD}), neglecting nonlinear terms and higher order
dispersion we obtain the linear part 
\begin{equation}
\left[ i\left( \frac{\partial }{\partial t}+v_{g}\frac{\partial }{\partial z}%
\right) +\frac{v_{g}^{\prime }}{2}\frac{\partial ^{2}}{\partial z^{2}}\right]
\widetilde{B}_{x}=0,
\end{equation}%
of the one-dimensional NLS equation. Here $v_{g}$ is the group velocity and $%
v_{g}^{\prime }=dv_{g}/dk$ is the group velocity dispersion. Since the spin
effects do not enter linear theory, these coefficients are the same in our
classical and quantum mechanical models.


\section{Classical Nonlinear Theory}



To explore nonlinearities due to relativistic effects, the momentum equation
is modified by letting $\mathbf{v}_{s}\rightarrow \gamma \mathbf{v}_{s}$ in
the left hand side, and it thus reads 
\begin{equation}
n_{s}\left( \frac{\partial }{\partial t}+\mathbf{v}_{s}\cdot \nabla \right)
\gamma \mathbf{v}_{s}=\frac{q_{s}}{m_{s}}n_{s}\left( \mathbf{E}+\mathbf{v}%
_{s}\times \mathbf{B}\right) -v_{s\mathrm{t}}^{2}\nabla n_{s},  \label{rel}
\end{equation}%
where $\gamma =1/(1-v^{2}/c^{2})$ and we have introduced the thermal
velocity $v_{s\mathrm{t}}=(k_{B}T/m_{s})^{1/2}$ for species $s$. The $\gamma 
$-factor can be Taylor expanded to first order and will thus result in
purely cubic nonlinearity in the velocity. Including this nonlinearity only,
it is straightforward to deduce the NLS equation 
\begin{equation}
\left[ i\left( \frac{\partial }{\partial t}+v_{g}\frac{\partial }{\partial z}%
\right) +\frac{v_{g}^{\prime }}{2}\frac{\partial ^{2}}{\partial z^{2}}+\frac{%
2\alpha ^{3}B_{0}k^{6}}{\omega _{p}^{2}\left( 1+{c^{2}k^{2}}/{\omega _{p}^{2}%
}\right) ^{2}}|\widetilde{B}_{x}|^{2}\right] \widetilde{B}_{x}=0.
\end{equation}%
The nonlinear term above will be complemented by nonlinear density
modifications induced by the ponderomotive force. Within a model that only
includes the electron dynamics, the density modifications will be limited
due to the general tendency of charge neutrality. However, for sufficiently
long pulses the low frequency ion dynamics will start to contribute to the
nonlinear behavior of the electrons, and it turns out that a fair comparison
between quantum and classical nonlinearities must include this effect. To
capture the ponderomotive nonlinearities, we start with Eq.\ (\ref{rel})
(for simplicity omitting the relativistic contribution, that we already
know), for a two-fluid ion-electron model 
. Again neglecting the displacement current in Amp\`{e}res law, linearizing
as previously and using the Maxwell equations we obtain the system 
\begin{equation}
\text{det}\left[ \hat{D}_{\mathrm{op}}\right] \left( 
\begin{array}{c}
v_{e\mathrm{lf}} \\ 
E_{\mathrm{lf}} \\ 
n_{e\mathrm{lf}} \\ 
v_{i\mathrm{lf}} \\ 
n_{i\mathrm{lf}}%
\end{array}%
\right) =\text{Adj}\left[ \hat{D}_{\mathrm{op}}\right] \left( 
\begin{array}{c}
\frac{-2}{\mu _{0}m_{e}}\frac{\partial }{\partial z}|\widetilde{B}_{x}|^{2}
\\ 
0 \\ 
0 \\ 
\frac{-2}{\mu _{0}m_{i}}\frac{\partial }{\partial z}|\widetilde{B}_{x}|^{2}
\\ 
0%
\end{array}%
\right) ,  \label{lflongmatrix}
\end{equation}%
for the low-frequency variables, where 
\begin{equation}
\hat{D}_{\mathrm{op}}=\left( 
\begin{array}{ccccc}
n_{0}\frac{\partial }{\partial t} & -\frac{q_{e}n_{0}}{m_{e}} & \frac{%
k_{b}T_{e}}{m_{e}}\frac{\partial }{\partial z} & 0 & 0 \\ 
0 & \frac{\partial }{\partial z} & -\frac{q_{e}}{\epsilon _{0}} & 0 & -\frac{%
q_{i}}{\epsilon _{0}} \\ 
n_{0}\frac{\partial }{\partial z} & 0 & \frac{\partial }{\partial t} & 0 & 0
\\ 
0 & -\frac{q_{i}n_{0}}{m_{i}} & 0 & n_{0}\frac{\partial }{\partial t} & 
\frac{k_{b}T_{i}}{m_{i}}\frac{\partial }{\partial z} \\ 
0 & 0 & 0 & n_{0}\frac{\partial }{\partial z} & \frac{\partial }{\partial t}
\\ 
&  &  &  & 
\end{array}%
\right) ,
\end{equation}%
and we can read off that 
\begin{eqnarray}
v_{e\mathrm{lf}} &=&\kappa _{e}|\widetilde{B}_{x}|^{2}, \\
v_{i\mathrm{lf}} &=&\kappa _{i}|\widetilde{B}_{x}|^{2}.
\end{eqnarray}%
The coefficients $\kappa _{e}$ and $\kappa _{i}$ are determined by solving
the corresponding differential equation, obtained from Eq. (\ref%
{lflongmatrix}), using Greens function techniques, and the result is: 
\begin{equation}
\kappa _{e}\approx -2v_{g}\frac{{\omega _{pi}^{2}}/{m_{e}}+{\omega _{pe}^{2}}%
/{m_{i}}}{n_{0}\mu _{0}\left[ \omega _{pe}^{2}(v_{g}^{2}-v_{it}^{2})+\omega
_{pi}^{2}(v_{g}^{2}-v_{et}^{2})\right] }.
\end{equation}

Now that the low frequency perturbations have been determined, the back
reaction on the original time scale can be calculated. We note that on this
fast time scale we return to neglecting ion motion. Then, we obtain 
\begin{equation}
\begin{split}
& \nabla \times \left[ n_{0}v_{e\mathrm{lf}}\frac{\partial }{\partial z}%
\left( \frac{1}{\mu _{0}q_{e}n_{0}}\frac{\partial }{\partial z}\hat{\mathbf{z%
}}\times \mathbf{B}\right) +\frac{q_{e}}{m_{e}}n_{e\mathrm{lf}}\mathbf{v}%
\times \mathbf{B}_{0}\right] _{x\,} \\
& \quad =i\left( \frac{k^{3}}{\mu _{0}q_{e}}-\frac{k^{2}B_{0}}{m_{e}\mu
_{0}v_{g}}\right) \kappa _{e}|\widetilde{B}_{x}|^{2}\widetilde{B}%
_{x}e^{i\beta },
\end{split}%
\end{equation}%
where the subscript $x$ indicates the $\hat{x}$-component of the vector.
Combining this result with the linear theory we obtain an NLS equation that
reads 
\begin{equation}
\begin{split}
& i\left( \frac{\partial }{\partial t}+v_{g}\frac{\partial }{\partial z}%
\right) \widetilde{B}_{x}+\frac{v_{g}^{\prime }}{2}\frac{\partial ^{2}%
\widetilde{B}_{x}}{\partial z^{2}} \\
& \qquad +\left[ \frac{{q_{e}}B_{0}k^{2}c^{2}/{m_{e}v_{g}}-k^{3}c^{2}}{%
\omega _{pe}^{2}\left( 1+{c^{2}k^{2}}/{\omega _{pe}^{2}}\right) }\kappa _{e}%
\right] |\widetilde{B}_{x}|^{2}\widetilde{B}_{x}=0.
\end{split}
\label{eq:nlse1}
\end{equation}


\section{Fully nonlinear theory}


As opposed to the case of a normal EMHD plasma, in the quantum case we have
one equation for each electron spin direction, and the extra term $\pm \mu
_{B}n_{\pm }\nabla B$ due to the spins influence on the magnetization. One
can note that if the spin populations are exactly equal in density, when
adding the two force equations this term will vanish, and this corresponds
to the classical case. However, if there is a slight difference in number
density, nonlinear fluctuations in the magnetic field will be induced. To
explore this effect, we try to derive an EMHD model , but now using the
two-fluid spin model. Eq. (\ref{EMHD}) is then replaced by 
\begin{eqnarray}
2Nq_{e}\frac{\partial \mathbf{B}}{\partial t} &=&\frac{1}{\mu _{0}}\nabla
\times \left[ \left( \nabla \times \mathbf{B}\right) \times \mathbf{B}\right]
-2\mu _{B}\nabla \times \left[ \left( \nabla \times (n\hat{\mathbf{B}}
)\right) \times \mathbf{B}\right]
\notag \\
&&
+2\mu _{B}\nabla \times \left( n\nabla
B\right) ,  \label{spinsum}
\end{eqnarray}
where the average number of electrons $N=(n_{+}+n_{-})/2$ tend to be deviate
little from the unperturbed density (due to charge neutrality), but the
difference between the electron species $n=(n_{+}-n_{-})/2$ can vary more.
Furthermore we introduce the notation $\mathbf{V}=\mathbf{v}_{+}+\mathbf{v}%
_{-}$ and $\mathbf{v}=\mathbf{v}_{+}-\mathbf{v}_{-}$. We here point out that
in the approximation considered, where the unperturbed density difference
difference is neglected, similar as before the linear treatment give
agreement with the classical case.

Next, to calculate the low frequency perturbations of $n$, we need the
difference between Eqs. (\ref{spinmomentum}) for the two species, which is
written 
\begin{equation}
\begin{split}
& 0=-\left[ \left( N\mathbf{v}+n\mathbf{V}\right) \times \mathbf{B}\right] \\
& \quad -2N\frac{\mu _{B}}{q_{e}B_{0}}\frac{\partial }{\partial z}\left(
B_{1x}^{2}+B_{1y}^{2}\right) +\frac{2k_{B}T}{q_{e}}\frac{\partial n}{%
\partial z}.
\end{split}%
\end{equation}%
Filtering out the low frequency time scale we obtain: 
\begin{equation}
2\left( \frac{\partial ^{2}}{\partial z^{2}}-\frac{\omega _{p}^{2}}{v_{\text{%
te}}^{2}}\right) N_{\text{lf}}=-\frac{1}{\mu _{0}mv_{\text{te}}^{2}}\frac{%
\partial ^{2}}{\partial z^{2}}\left( |\widetilde{B}_{x}|^{2}+|\widetilde{B}%
_{y}|^{2}\right) ,
\end{equation}%
and 
\begin{equation}
n_{\text{lf}}=2N\frac{\mu _{B}}{k_{B}TB_{0}}\left( |\widetilde{B}_{x}|^{2}+|%
\widetilde{B}_{y}|^{2}\right)
\end{equation}%
where we have introduced the electron thermal velocity, $v_{\text{te}%
}=(k_{B}T/m_{e})^{1/2}$. Due to the reduced geometry of the problem, with
parallel propagating circularly polarized modes, all second harmonic density
perturbations can be neglected, and all second order nonlinearities in the
magnetic field also vanish. Furthermore, $N_{\mathrm{lf}}\ll n_{\mathrm{lf}}$
as a consequence of the system tending towards charge neutrality, and thus
only $n_{\mathrm{lf}}$ needs to be considered here.

Inserting this now in Eq. (\ref{spinsum}) and considering the original time
scale the only nonvanishing contribution to the nonlinear constant is 
\begin{equation}
-\mu _{B}\nabla \times \left[ \left( \nabla \times (n\hat{\mathbf{B}}%
)\right) \times \mathbf{B}\right] _{x\,\text{ }}=16iN\frac{\mu _{B}^{2}k^{2}%
}{k_{B}TB_{0}}|\widetilde{B}_{x}|^{2}\widetilde{B}_{x}.
\end{equation}%
Thus, the full NLS equation including all the effects discussed above
(relativistic nonlinearity, classical density perturbations induced by the
ponderomotive force, and a spin dependent density modification, driven by
the nonlinear magnetic dipole force) will be (cf.\ \ Eq. (\ref{eq:nlse1})) 
\begin{equation}
\begin{split}
& i\left( \frac{\partial }{\partial t}+v_{g}\frac{\partial }{\partial z}%
\right) \widetilde{B}_{x}+\frac{v_{g}^{\prime }}{2}\frac{\partial ^{2}%
\widetilde{B}_{x}}{\partial z^{2}}+\frac{4\mu _{B}^{2}k^{2}}{k_{B}TB_{0}q_{e}%
}|\widetilde{B}_{x}|^{2}\widetilde{B}_{x} \\
& +\left[ \frac{2\alpha _{e}^{3}B_{0}k^{6}}{\omega _{pe}^{2}\left( 1+{%
c^{2}k^{2}}/{\omega _{pe}^{2}}\right) ^{2}}+\frac{\frac{q_{e}}{m_{e}v_{g}}%
B_{0}k^{2}c^{2}-k^{3}c^{2}}{\omega _{pe}^{2}\left( 1+{c^{2}k^{2}}/{\omega
_{pe}^{2}}\right) }\kappa _{e}\right] |\widetilde{B}_{x}|^{2}\widetilde{B}%
_{x}=0.
\end{split}
\label{eq:nlse2}
\end{equation}%
This equation is the main result of this paper. From the magnitude of the
nonlinear coefficient, one can determine the regimes in which the spin terms
can dominate and be responsible for e.g. soliton formation.


\section{Discussion}


In the present paper we have studied weakly nonlinear whistler waves
propagating along the magnetic field. A nonlinear Schr\"{o}dinger equation
has been derived for the case of classical nonlinerities (see Eq.~(\ref%
{eq:nlse2})). The nonlinear coefficient than gets two contributions; from
relativistic effects and from low-frequency density modifications induced by
the ponderomotive force. Taking spin effects into account, within in a
electron two-fluid spin model, it is found that the low-frequency part of
the magnetic dipole force separates the spin up and spin down populations.
Due to the different magnetization currents from the two populations, a spin
contribution to the nonlinear coefficient then arises. Firstly, comparing
the spin contribution to the nonlinear coefficient with the relativistic
contribution, we see that the former is larger provided that 
\begin{equation}
1\lesssim \frac{m_{i}\hbar ^{2}\omega _{pe}^{2}}{k_{B}Tm_{e}^{2}C_{A}^{2}}
\label{Eq:rel-cond}
\end{equation}%
Here we have used the lowest frequency allowed by the model $\omega \sim
\omega _{ci}$ to get a condition that is relatively easy to fulfill.
However, we must also compare the spin induced nonlinearity against the
contribution from the nonlinear density oscillations induced by the
ponderomotive force. It is found that the former dominates when 
\begin{equation}
1\lesssim \frac{\hbar \omega _{ce}}{m_{i}C_{A}^{2}}\frac{\hbar \omega _{ce}}{%
m_{e}v_{it}^{2}}  \label{Eq:Pond-compare}
\end{equation}%
where we have used that the maximum value of $kc$ is roughly $\omega _{pe}$,
due to the limitations imposed by the geometry in combination with the EMHD
approximation. The factor $\hbar \omega _{ce}/m_{i}C_{A}^{2}$is the
condition for nonlinear spin effects to dominate, when a similar comparison
is made in the standard MHD regime, according to Ref. \cite{Brodin-2008}. To
get a more favorable comparison (i.e. a condition that is easier to reach
under laboratory conditions) than in these previous works, the second
factor, $\hbar \omega _{ce}/m_{i}v_{ti}^{2}$, must be larger than unity.
This is unfortunately not the case for the parameters usually found in
laboratory conditions. However, astrophysical plasmas with parameters
fulfilling both conditions (\ref{Eq:rel-cond}) and (\ref{Eq:Pond-compare})
can be found, e.g. in the vicinity of pulsars or magnetars \cite{Astro}, and
thus we note that effects associated with the electron spin can be more
important than the classical relativistic and ponderomotive nonlinearities
in such environments.

The present study has focused on the EMHD regime. While it is shown that
spin effect certainly can be important during e.g. astrophysical plasma
conditions, our study suggest the standard MHD regime \cite{Brodin-2008} can
be more affected by the electron spin properties during laboratory
conditions . However, much more work remains to be done in order for this
conclusion to be settled, as the picture may change when a more general
geometry is considered, or when kinetic effects \cite{g-factor} are taken
into account.

\acknowledgments
This research is supported by the European Research Council under Constract
No.\ 204059-QPQV and the Swedish Research Council under Contracts No.\
2007-4422.


\begin{thebibliography}{99}
\bibitem{Bohm-Pines} D. Bohm and D. Pines, Phys. Rev. \textbf{92}, 609
(1953).

\bibitem{Pines-1953} D. Pines, Phys. Rev. \textbf{92}, 626 (1953).

\bibitem{Haas-2000} F. Haas, G. Manfredi, and M. R. Feix, Phys. Rev. E 
\textbf{62}, 2763 (2000).

\bibitem{Manfredi-review} G. Manfredi, Fields Inst. Comm. \textbf{46}, 263
(2005).

\bibitem{Garcia-2005} L. G. Garcia, F. Haas, L. P. L. de Oliviera, and J.
Goedert, Phys. Plasmas \textbf{12}, 012302 (2005).

\bibitem{Shukla-Stenflo-2006} P. K. Shukla and L. Stenflo, Phys. Lett. A 
\textbf{355}, 378 (2006).

\bibitem{Shukla-Eliasson-2006} P. K. Shukla and B. Eliasson, Phys. Rev.
Lett. \textbf{96}, 245001 (2006).

\bibitem{Marklund-2007} M. Marklund and G. Brodin, Phys. Rev. Lett. \textbf{%
98}, 025001 (2007).

\bibitem{Brodin-2007} G. Brodin and M. Marklund, New J. Phys. \textbf{9},
277 (2007).

\bibitem{Shukla-2007} P. K. Shukla and B. Eliasson, Phys. Rev. Lett. \textbf{%
99}, 096401 (2007).

\bibitem{Brodin-2008} G. Brodin, M. Marklund and G. Manfredi, Phys. Rev.
Lett. \textbf{100}, 175001 (2008)

\bibitem{g-factor} G. Brodin, M. Marklund, J. Zamanian, \AA . Ericsson, and
P.L. Mana, \textbf{101,} 245002 (2008).

\bibitem{Atwater-Plasmonics} H. A. Atwater, Sci. Am.\textbf{\ 296}, 56
(2007).

\bibitem{Marklund-EPL-plasmonics} M. Marklund, G. Brodin, L. Stenflo, and C.
S. Liu, Europhys. Lett. \textbf{84},17006 (2008).

\bibitem{Manfredi-quantum-well} G. Manfredi and P.-A. Hervieux, Appl. Phys.
Lett. \textbf{91}, 061108 (2007).

\bibitem{Ultracold} M. P. Robinson, B. Laburthe Tolra, M. W. Noel, et al.,
Phys. Rev. Lett. \textbf{85}, 4466 (2000).

\bibitem{Astro} A. K. Harding and D. Lai, Rep. Prog. Phys. \textbf{69}, 2631
(2006).

\bibitem{Melrose-book} D. B. Melrose, \textit{Quantum Plasmadynamics}
(Springer-Verlag, 2008).

\bibitem{Cowley-1986} S. C. Cowley, R. M. Kulsrud, and E. Valeo, Phys.
Fluids \textbf{29}, 430 (1986).

\bibitem{Kulsrud-1986} R. M. Kulsrud, E. J. Valeo, and S. C. Cowley, Nucl.
Fusion \textbf{26}, 1443 (1986).

\bibitem{NJP1} G. Brodin and M. Marklund, New J. Phys. \textbf{10}, 115031
(2008).

\bibitem{NJP2} J. Zamanian, G. Brodin, and M. Marklund, New J. Phys. \textbf{%
11}, 073017 (2009).

\bibitem{AIPConf} G. Brodin, M. Marklund, and J. Zamanian, AIP Conf. Proc. 
\textbf{1041}, 97 (2008).

\bibitem{shukla-ali-etal.} P. K. Shukla, S. Ali, L. Stenflo, and M.
Marklund, Phys. Plasmas \textbf{13}, 112111 (2006).

\bibitem{nitin} N. Shukla, G. Brodin, M. Marklund, P. K. Shukla, and L.
Stenflo, Phys. Plasmas \textbf{16}, 072114 (2009).

\bibitem{shukla-eliasson-review} P. K. Shukla and B. Eliasson,
arXiv:0906.4051 (2009).

\bibitem{Melrose-2009} D. B. Melrose and A. Mushtaq, Phys. Plasmas \textbf{16%
}, 094508 (2009).

\bibitem{bulanov91} S.V. Bulanov, F. Pegoraro, A. S. Sakharov, Magnetic
reconnection in electron magnetohydrodynamics, 1992.
\end{thebibliography}
\end{document}